%% file: arxiv_IV.tex
\documentclass[numbers,sort&compress]{article}
\usepackage{graphicx}
\usepackage{listings}
\usepackage{url}
\usepackage{lineno}
\usepackage{subcaption}
\usepackage{rotating}
\usepackage[normalem]{ulem}
\usepackage{arxiv}
\usepackage{amssymb,amsmath,amsthm}
\usepackage[numbers,sort&compress]{natbib}
\usepackage{xcolor}
\usepackage{hyperref}
\usepackage{tabularx}
\usepackage[toc,page]{appendix}
\graphicspath{{Artworks/}{imgs/}}
\usepackage{algorithm2e}
\theoremstyle{definition}
\usepackage{amsfonts}       %
\usepackage{nicefrac}       %
\usepackage{microtype}      %
\usepackage{cleveref}       %
\usepackage{doi}
\usepackage{tikz}
\usetikzlibrary{positioning,arrows.meta,fit,backgrounds,calc,shapes.geometric}
\definecolor{RuhiTeal}{HTML}{004D40}
\definecolor{RuhiCoral}{HTML}{FF655D}
\definecolor{RuhiYellow}{HTML}{F1DB4B}
\input{authors_arxiv.tex}
\date{\today}
\title{Two-dimensional RMSD projections for reaction path visualization and validation}
\hypersetup{
 pdfauthor={},
 pdftitle={Two-dimensional RMSD projections for reaction path visualization and validation},
 pdfkeywords={},
 pdfsubject={},
 pdfcreator={Emacs 30.2 (Org mode 9.7.39 + HaoZeke)}, 
 pdflang={English}}
\begin{document}

\maketitle
\begin{abstract} %
Transition state or minimum energy path finding methods constitute a routine
component of the computational chemistry toolkit. Standard analysis involves
trajectories conventionally plotted in terms of the relative energy to the
initial state against a cumulative displacement variable, or the image number.
These dimensional reductions obscure structural rearrangements in high
dimensions and are often history dependent. This precludes the ability to
compare optimization histories of different methods beyond the number of
calculations, time taken, and final saddle geometry. We present a method mapping
trajectories onto a two-dimensional projection defined by a permutation corrected
root mean square deviation from the reactant and product configurations. Energy
is represented as an interpolated color-mapped surface constructed from all
optimization steps using a gradient-enhanced Gaussian Process with the inverse
multiquadric kernel, whose posterior variance contours delineate data-supported
regions from extrapolated ones. A rotated coordinate frame decomposes the RMSD
plane into reaction progress and orthogonal distance. We show the utility of the
framework on a cycloaddition reaction, where a machine-learned potential
saddle and density functional theory reference lie on comparable energy contours
despite geometric displacements, along with the ratification of the
visualization for more complex reactions, a Grignard rearrangement, and a
conrotatory bicyclobutane ring opening.
\end{abstract}

The first order saddle point is often the first approximation to understand the
reaction kinetics of any system of interest. Starting from two known
configurations, a series of configurations connected through fictitious springs,
that is, the nudged elastic band method \cite{jonssonNudgedElasticBand1998} is
among the most common. The NEB may be coupled with the climbing image
\cite{henkelmanClimbingImageNudged2000} and spring variations
\cite{asgeirssonNudgedElasticBand2021} with machine learned local accelerations
\cite{goswamiEfficientImplementationGaussian2025a,koistinenNudgedElasticBand2019,goswamiEnhancedClimbingImage2026}
and the string family of methods
\cite{petersGrowingStringMethod2004,zimmermanGrowingStringMethod2013,marksIncorporationInternalCoordinates2025}.
Together, these constitute the current Pareto front for saddle searches.

Beyond convergence difficulties and the sensitivity to initial points,
the resulting path histories are best studied through the ``eye-ball
norm'', wherein manual inspection of the final path, along with a normal mode
analysis for the saddle point estimate are considered sufficient. In turn, this
means the most common visual plots are one dimensional ``profile'' plots. These
may show only the final optimized path, against the image number \footnote{e.g. as
found in ASE \cite{larsenAtomicSimulationEnvironment2017}}, which obscures any
distance measure between the images. Alternatively, a well-known improvement to
this involves the cubic Hermite interpolation
\cite{henkelmanClimbingImageNudged2000} involving the forces relative to the
``reaction coordinate,'' defined by the piece-wise sum of the Euclidean distances
between intermediate images. The reaction coordinate \(s_i = \sum_{j=1}^{i} \|
\mathbf{R}_j - \mathbf{R}_{j-1} \|_2\) depends entirely on the specific path
geometry and optimization history. This scalar measure lacks a unique definition
and collapses global geometric information onto a single, arbitrary axis. This
dimensional reduction precludes rigorous comparison between trajectories
generated by differing algorithms, such as NEB versus Frozen String methods \cite{marksIncorporationInternalCoordinates2025}, or
even identical algorithms with varying parameters. The
one-dimensional projection frequently masks the distinction between numerical
instability and physical relaxation into alternative basins. The resulting
ambiguity in stationary point validation costs researchers days of
calculations to clarify. 
\section{Methodology}
\label{sec:org360bc3f}

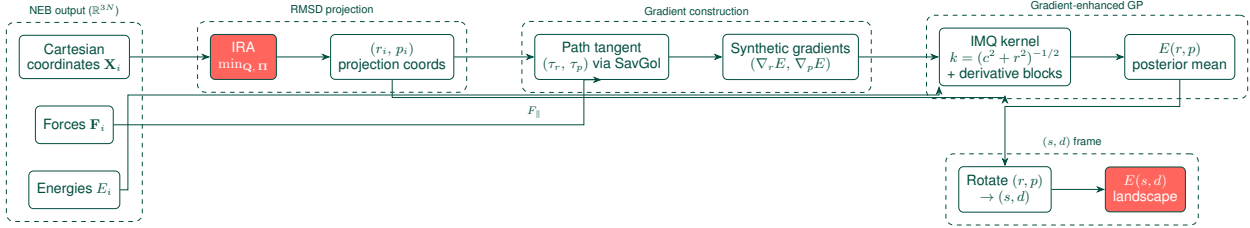
\begin{figure*}[t!]
\centering
\resizebox{\textwidth}{!}{%
\begin{tikzpicture}[
  >=Stealth,
  node distance=6mm and 12mm,
  box/.style={draw=RuhiTeal, fill=white, rounded corners=3pt,
              text=RuhiTeal, minimum height=2.4em, align=center,
              font=\small\sffamily, inner sep=5pt},
  coral/.style={box, fill=RuhiCoral, text=white},
  grp/.style={draw=RuhiTeal, dashed, rounded corners=5pt, inner sep=8pt},
  lbl/.style={font=\scriptsize\sffamily, text=RuhiTeal},
  arr/.style={->, RuhiTeal, semithick},
]
\node[box] (Xi) {Cartesian\\coordinates $\mathbf{X}_i$};
\node[box, below=of Xi] (Fi) {Forces $\mathbf{F}_i$};
\node[box, below=of Fi] (Ei) {Energies $E_i$};
\begin{scope}[on background layer]
  \node[grp, fit=(Xi)(Fi)(Ei), label={[lbl]above:NEB output ($\mathbb{R}^{3N}$)}] (gin) {};
\end{scope}

\node[coral, right=18mm of Xi] (ira) {IRA\\$\min_{\mathbf{Q},\,\boldsymbol{\Pi}}$};
\node[box, right=of ira] (rp) {$(r_i,\, p_i)$\\projection coords};
\begin{scope}[on background layer]
  \node[grp, fit=(ira)(rp), label={[lbl]above:RMSD projection}] (grmsd) {};
\end{scope}

\node[box, right=18mm of rp] (tan) {Path tangent\\$(\tau_r,\,\tau_p)$ via SavGol};
\node[box, right=of tan] (syn) {Synthetic gradients\\$(\nabla_r E,\,\nabla_p E)$};
\begin{scope}[on background layer]
  \node[grp, fit=(tan)(syn), label={[lbl]above:Gradient construction}] (ggrad) {};
\end{scope}

\node[box, right=18mm of syn] (imq) {IMQ kernel\\$k=(c^2+r^2)^{-1/2}$\\+ derivative blocks};
\node[box, right=of imq] (surf) {$E(r,p)$\\posterior mean};
\begin{scope}[on background layer]
  \node[grp, fit=(imq)(surf), label={[lbl]above:Gradient-enhanced GP}] (ggp) {};
\end{scope}

\node[box, below=18mm of imq] (rot) {Rotate $(r,p)$\\$\to (s,d)$};
\node[coral, right=of rot] (sd) {$E(s,d)$\\landscape};
\begin{scope}[on background layer]
  \node[grp, fit=(rot)(sd), label={[lbl]above:$(s,d)$ frame}] (gsd) {};
\end{scope}

\draw[arr] (Xi) -- (ira);
\draw[arr] (ira) -- (rp);
\draw[arr] (Fi) -| ([xshift=-4mm]tan.south) -- (tan.south);
\draw[arr] (rp) -- (tan);
\draw[arr] (tan) -- (syn);
\draw[arr] (syn) -- (imq);
\draw[arr] (imq) -- (surf);
\draw[arr] (surf.south) -- ++(0,-6mm) -| (rot);
\draw[arr] (rot) -- (sd);
\draw[arr] (Ei) -- ++(0:12mm) |- ([yshift=-2mm]imq.south west) -- (imq.south west);
\draw[arr] (rp.south) -- ++(0,-4mm) -| ([yshift=-2mm]imq.south);

\node[lbl, above=1pt of Fi -| tan.south west] {$F_\parallel$};
\end{tikzpicture}%
}
\caption{Overview of the 2D RMSD projection pipeline. NEB output in
$\mathbb{R}^{3N}$ is projected to intrinsic $(r, p)$ coordinates via
IRA-based permutation-invariant RMSD\@. Tangential forces are projected onto
the 2D tangent to construct synthetic gradients. Energies and gradients jointly
feed a gradient-enhanced GP with the IMQ kernel to produce the interpolated
energy surface $E(r,p)$. A final rigid rotation aligns the $(r, p)$ frame to
reaction progress $s$ and orthogonal deviation $d$, yielding the $E(s,d)$
landscape.}
\label{fig:pipeline}
\end{figure*}

The complete methodology is summarized in Figure \ref{fig:pipeline}. We describe the
one-dimensional profiles first. Standard NEB implementations provide
discrete images \(\mathbf{X}_i\) and energies \(E_i\). To reconstruct a physically
consistent energy profile \(E(s)\) along the path, we use the available force
information. We define the discrete reaction coordinate \(s_i\) as the cumulative
Euclidean distance:

\begin{equation}
  s_i = \sum_{j=1}^{i} \| \mathbf{X}_j - \mathbf{X}_{j-1} \|
\end{equation}

Instead of simple linear or cubic interpolation, we employ a Piecewise Cubic
Hermite Interpolating Polynomial (PCHIP). This method utilizes the tangent
forces \(F_{\parallel, i}\) to constrain the derivative of the energy surface:

\begin{equation}
  \frac{dE}{ds}\bigg|_{s_i} = -F_{\parallel, i} = -(\mathbf{F}_i \cdot \hat{\tau}_i)
\end{equation}

where \(\hat{\tau}_i\) denotes the unit tangent vector along the path.
\subsection{Intrinsic Projection Coordinates}
\label{sec:org59b564a}
To map the high-dimensional optimization trajectory onto a consistent 2D
subspace, we define the projection coordinates \((r, p)\) as the distances from
the reactant (\(\mathbf{R}\)) and product (\(\mathbf{P}\)) reference configurations.
To ensure the metric remains robust across automated workflows where atom
indexing may vary, we employ a permutation-invariant Root Mean Square Deviation
(RMSD) with optimal permutation \(\boldsymbol{\Pi}\) and rotation \(\mathbf{Q}\)
determined via the Iterative Rotations and Assignments (IRA) 
\cite{gundeIRAShapeMatching2021} algorithm. This procedure guarantees unique,
invariant coordinates regardless of the initial atom indexing or frame
orientation \cite{goswamiEnhancedClimbingImage2026}. The discrete optimization
steps provide a sparse sampling of this geometric subspace.

For a system with \(N\) atoms having positions \(\mathbf{X} \in \mathbb{R}^{3
\times N}\), we define the distance metric \(d(\mathbf{X},
\mathbf{X}_{\text{ref}})\) as:

\begin{equation}
  d(\mathbf{X}, \mathbf{X}_{\text{ref}}) = \min_{\mathbf{Q}, \mathbf{\Pi}} \sqrt{ \frac{1}{N} \left\| \mathbf{X} - \mathbf{Q} \mathbf{X}_{\text{ref}} \mathbf{\Pi} \right\|_F^2 }
\end{equation}

Here, \(\|\cdot\|_F\) denotes the Frobenius norm. \(\mathbf{Q} \in SO(3)\)
represents the optimal rotation matrix, and \(\mathbf{\Pi}\) represents the
optimal permutation matrix. We solve for \(\mathbf{Q}\) and \(\mathbf{\Pi}\)
simultaneously. This ensures that the resulting coordinates \((r, p) =
(d(\mathbf{X}, \mathbf{R}), d(\mathbf{X}, \mathbf{P}))\) remain invariant to
rigid body rotation, translation, and arbitrary atom index labeling.
\subsection{Reaction Progress Coordinates}
\label{sec:org8b66af2}

The raw \((r, p)\) plane contains an unphysical region where both RMSD distances
are simultaneously small, which has no counterpart in configuration space for
non-trivial rearrangements. We apply a rigid rotation that decomposes the RMSD
plane into reaction progress \(s\) and orthogonal deviation \(d\), analogous to the
path collective variable decomposition 
\cite{branduardiFreeEnergySpace2007}. Given the first and last path images
with projected coordinates \((r_0, p_0)\) and \((r_N, p_N)\), we define the unit
tangent \((\hat{s}_r, \hat{s}_p)\) along the path direction and the normal
\((\hat{d}_r, \hat{d}_p) = (-\hat{s}_p, \hat{s}_r)\). The rotated coordinates
are then:

\begin{align}
s_i &= (r_i - r_0)\,\hat{s}_r + (p_i - p_0)\,\hat{s}_p \\
d_i &= (r_i - r_0)\,\hat{d}_r + (p_i - p_0)\,\hat{d}_p
\end{align}

where \(s\) measures progress along the reaction path and \(d\) measures
perpendicular deviation from it, both in Angstroms. The free energy
literature \cite{branduardiFreeEnergySpace2007} defines path progress and orthogonal
distance via an exponential softmin over \(N\) reference frames, \(s =
\sum_i i\,e^{-\lambda R_i} / \sum_i e^{-\lambda R_i}\) and \(z = -(1/\lambda)
\ln \sum_i e^{-\lambda R_i}\), with a smoothing parameter \(\lambda\). That
formulation targets enhanced sampling methods (metadynamics, umbrella sampling)
where smooth, differentiable collective variables are required for biasing
forces at finite temperature. For post-hoc visualization of a zero-temperature
discrete path, the linear rotation suffices as it is parameter-free, exact at the
endpoints, and directly invertible. Both decompositions separate progress along
a path from deviation orthogonal to it. All figures in this work use this
\((s, d)\) frame.

Unlike principal component analysis (PCA)
\cite{jamesIntroductionStatisticalLearning2013} or t-distributed stochastic
neighbor embedding (t-SNE) \cite{maatenVisualizingDataUsing2008} applied to
covariance matrices
\cite{musaelianLearningLocalEquivariant2023,ceriottiUnsupervisedMachineLearning2019},
which require a priori selection of descriptors (e.g., bond lengths, angles,
SOAP vectors \cite{caroOptimizingManybodyAtomic2019}), RMSD-based projection
operates directly on Cartesian coordinates without feature engineering.
PCA-based methods popular in machine learning interatomic potential (MLIP)
communities presuppose that the dominant variance directions align with
chemically meaningful coordinates, an assumption that frequently fails for
complex rearrangements involving concerted bond breaking and formation.
Similarly, manifold learning techniques like t-SNE and UMAP
\cite{mcinnesUMAPUniformManifold2018} optimize for local neighborhood preservation
but lack the absolute geometric reference frame necessary for quantitative
cross-method comparison. Sketchmap
\cite{tribelloUsingSketchmapCoordinates2012} attempts to preserve both local and
global distance structure, but requires on the order of \(10^4-10^5\) samples
to learn a reliable mapping, and even then does not guarantee metric
preservation of absolute distances. A single NEB calculation produces \(\approx
10^3\) geometries at most. Our endpoint-distance coordinates provide a universal,
reaction-agnostic metric from exactly these \(10^3\) points, with no training
phase, no descriptor selection, and no presupposed chemical intuition about the
reaction mechanism.

For systems with well-established collective variables (Ramachandran angles for
peptides, donor-acceptor distances for proton transfer), such coordinates
provide more direct physical insight
\cite{mandelliModifiedNudgedElastic2021}. However, the vast majority of
transition state searches lack such convenient coordinates. Surface reactions,
where an adatom hops into a subsurface site on a metallic slab, or catalytic
processes involving concerted rearrangements of adsorbates, have no obvious
low-dimensional collective variable. Constructing a two-dimensional free
energy surface for such systems requires extensive sampling and a priori
knowledge of the relevant degrees of freedom. In contrast, the RMSD-based
projection requires only the Cartesian coordinates already produced by the
calculation. Unlike free energy surface methods that mandate pre-specification
of reaction coordinates, our approach visualizes arbitrary path based methods
post hoc, requiring only the endpoints. This eliminates the circular dependency
where one must already understand the mechanism to choose coordinates capable
of revealing it.

Our method targets the complementary regime to global PES exploration
techniques. It provides geometric validation for individual optimization
trajectories (\(\approx 10^3\) samples per NEB calculation) rather than exhaustive
basin catalogs requiring \(\approx 10^6-10^9\) samples. Most single NEB
calculations involve only \(\approx 10^3\) geometries concentrated along a
one-dimensional path connecting a single reactant-product pair. This sampling is
insufficient and inappropriate for techniques designed to map complete basin
connectivity or construct disconnectivity graphs
\cite{walesExploringEnergyLandscapes2018}. The framework instead asks whether this specific optimization found a physically
reasonable barrier topology. For MLIP validation, where the goal is verifying that a
machine-learned potential reproduces the correct transition pathway rather than
exhaustively cataloging all possible pathways, this geometric diagnostic
fills a gap that scalar comparisons cannot.
\subsection{Energy Landscape Projection}
\label{sec:org2492788}

\begin{figure}[t!]
\centering
\resizebox{\columnwidth}{!}{%
\begin{tikzpicture}[
  >=Stealth,
  node distance=8mm and 20mm,
  box/.style={draw=RuhiTeal, fill=white, rounded corners=3pt,
              text=RuhiTeal, minimum height=2.2em, align=center,
              font=\small\sffamily, inner sep=5pt},
  coral/.style={box, fill=RuhiCoral, text=white},
  note/.style={draw=RuhiTeal, fill=RuhiYellow, text=RuhiTeal,
               rounded corners=2pt, align=center, font=\scriptsize\sffamily,
               inner sep=4pt},
  grp/.style={draw=RuhiTeal, dashed, rounded corners=5pt, inner sep=10pt},
  lbl/.style={font=\scriptsize\sffamily, text=RuhiTeal},
  arr/.style={->, RuhiTeal, semithick},
]
\node[box, shape=ellipse] (pes) {PES: $E(\mathbf{R})$\\$\mathbf{R}\in\mathbb{R}^{3N-1}$};
\node[note, right=12mm of pes] (cfg)
  {Many configurations\\can share the same\\projected coordinates};
\begin{scope}[on background layer]
  \node[grp, fit=(pes)(cfg), label={[lbl]above:Configuration space}] {};
\end{scope}

\node[box, below left=30mm and 30mm of pes] (s1d)
  {$s$ = cumulative distance\\(path-dependent scalar)};
\node[coral, below=of s1d] (e1d) {$E(s)$};
\node[note, below=6mm of e1d] (loss1)
  {Loses: geometry, basin topology,\\cross-method comparison};
\begin{scope}[on background layer]
  \node[grp, fit=(s1d)(e1d)(loss1),
        label={[lbl]above:Standard: 1D profile}] {};
\end{scope}

\node[box, below right=30mm and 30mm of pes] (rp)
  {$(r,p)$ = RMSD to endpoints\\(intrinsic, path-independent)};
\node[box, below left=10mm and -4mm of rp] (rot)
  {Rotate $(r,p) \to (s,d)$\\progress + deviation};
\node[coral, right=of rot] (e2d) {$E(s,d)$};
\node[note, below=6mm of rot.south east] (loss2)
  {Loses: $3N{-}3$ dimensions\\Retains: endpoint distances,\\relative geometry};
\begin{scope}[on background layer]
  \node[grp, fit=(rp)(rot)(e2d)(loss2),
        label={[lbl]above:This work: 2D projection}] {};
\end{scope}

\draw[arr] (pes) -- node[lbl, left, align=center, pos=0.45]
  {project onto\\cumulative path length\\($3N{-}1\to1$)} (s1d);
\draw[arr] (pes) -- node[lbl, right, align=center, pos=0.45]
  {project onto\\RMSD distances to $R$, $P$\\($3N{-}1\to2$)} (rp);
\draw[arr] (s1d) -- (e1d);
\draw[arr] (rp) |- (rot);
\draw[arr] (rot) -- (e2d);
\end{tikzpicture}%
}
\caption{Dimensional reduction from the full $3N{-}1$ configuration space to 1D
(standard reaction coordinate $s$, left) versus 2D RMSD projection (this work,
right). Both maps are many-to-one, and distinct Cartesian configurations can share
the same projected coordinates. The 1D projection loses all geometric context.
The 2D projection retains endpoint distances for cross-method comparison
and landscape visualization, at the cost of $3N{-}3$ unconstrained degrees of
freedom.}
\label{fig:dimreduce}
\end{figure}
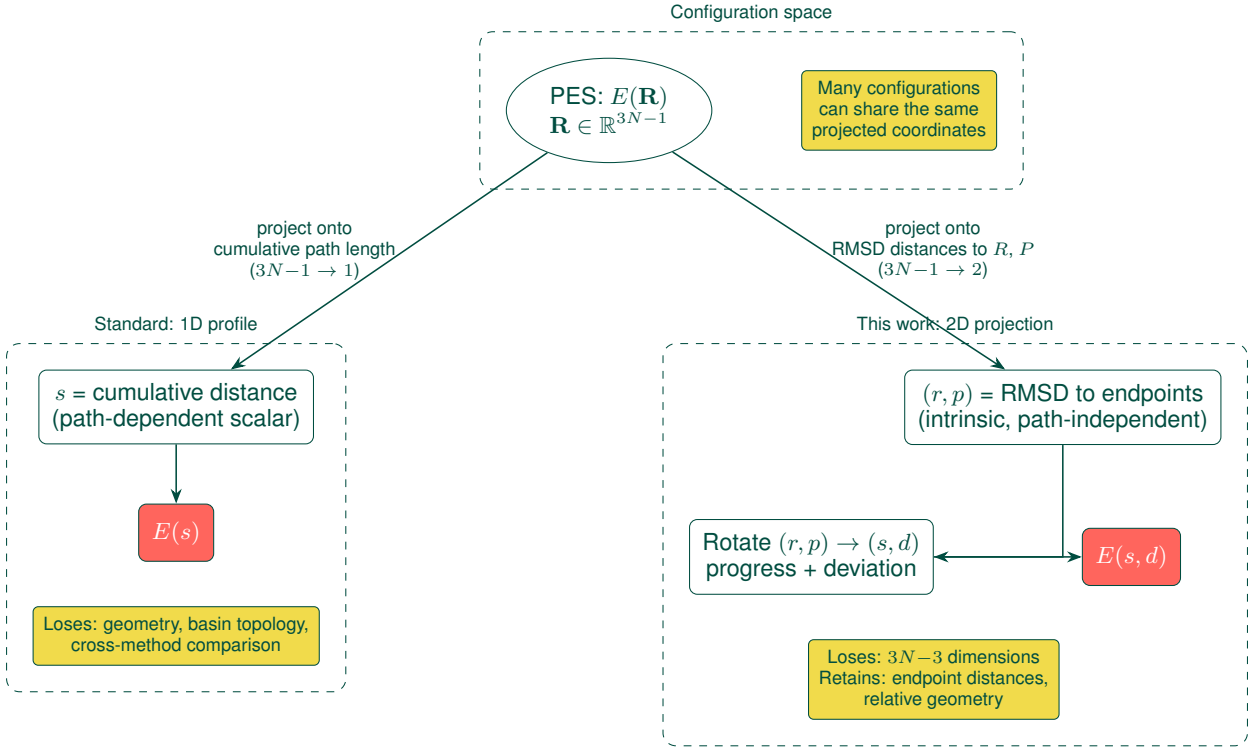

The discrete optimization steps provide a sparse, unstructured sampling of the
2D \((s, d)\) domain (Figure \ref{fig:dimreduce}). To visualize slices of the
underlying potential energy landscape \(E(s, d)\), in a meaningful way to compare
data from disparate calculations, we construct a continuous projection
approximating the potential energy surface. Both the 1D and 2D projections are
lossy, since multiple distinct configurations in \(\mathbb{R}^{3N}\) can map to the
same projected coordinate. The 2D projection retains strictly more information
than the 1D reaction coordinate, but it is not and cannot be the full PES.

While forces \(F = -\nabla_R(E)\) are available from electronic structure
calculations in 3N-dimensional Cartesian space, direct projection via the chain
rule would require computing \(\partial d_{\text{RMSD}}/\partial R\), a quantity
complicated by the minimization over permutations and rotations implicit in the
RMSD definition. Instead, we construct synthetic gradients in the 2D projection
space by combining the path tangent vector with the available parallel force
component.

First, we stabilize the tangent calculation via Savitzky-Golay smoothing of the
RMSD coordinates themselves:

\begin{equation}
\tilde{r}_i, \tilde{p}_i = \text{SavGol}(r_i, p_i; \text{window}=5, \text{poly}=2)
\end{equation}

From the smoothed coordinates, we compute the path tangent:

\begin{equation}
\tau_r = \frac{d\tilde{r}}{ds}, \quad \tau_p = \frac{d\tilde{p}}{ds}
\end{equation}

where the tangent is normalized so \(||\tau|| = 1\). The projected gradient
components are then:

\begin{equation}
\nabla_r E = -F_\parallel \cdot \tau_r, \quad \nabla_p E = -F_\parallel \cdot \tau_p
\end{equation}

where \(F_\parallel\) is the force component parallel to the path (column 3 in an eOn
\texttt{.dat} file). This construction ensures gradient information respects the local
path geometry while remaining tractable to compute from standard NEB output.

Only the tangential force component \(F_\parallel\) enters this construction. The
full Cartesian force \(\mathbf{F}_i \in \mathbb{R}^{3N}\) contains orthogonal
components that encode local curvature of the true PES. Projecting these
into the \((r, p)\) subspace would require the Jacobian \(\partial d_{\text{RMSD}} /
\partial \mathbf{R}\), which is not analytically available. The RMSD involves a
joint minimization over rotations \(\mathbf{Q} \in SO(3)\) and permutations
\(\mathbf{\Pi}\), and the resulting map from \(\mathbb{R}^{3N}\) to \(\mathbb{R}^2\)
is piecewise smooth at best, with non-differentiable boundaries wherever the
optimal permutation assignment changes. This is not a software limitation but a
mathematical property of the permutation-invariant RMSD itself. The tangential
projection is therefore the consistent choice without additional approximations.

The omission of orthogonal force components means the interpolated surface
cannot reproduce the curvature perpendicular to the path. In the full
\$3N\$-dimensional PES, MEP images are by definition minima in the hyperplane
orthogonal to the path; the 2D projection has no mechanism to enforce or verify
this property. The tangential projection nonetheless carries non-trivial 2D
information because the tangent direction \((\tau_r, \tau_p)\) varies along the
path. Force information is distributed across both projected coordinates in a
way that scalar reaction coordinate plots cannot access.

The interpolated surface is a projected slice of the full
\$3N\$-dimensional PES. Many distinct Cartesian configurations map to the same
\((r, p)\) point. The interpolation represents the conditional expectation of the
energy given the RMSD coordinates, marginalizing over the unconstrained \(3N-3\)
degrees of freedom. Features of the projected surface, such as apparent
alternative pathways in the color map, do not necessarily correspond to physical
pathways in the full configuration space. Similarly, the property that MEP
images are minima in the orthogonal hyperplane holds in the full-dimensional PES
but has no analogue in a two-dimensional lossy projection. The interpolation is
reliable near the sampled data and degrades with distance, as with any
regression model.

We denote the 2D projection coordinates compactly as \(\mathbf{x}_i = (r_i,
p_i) = (d_{\text{RMSD}}(S_i, R),\, d_{\text{RMSD}}(S_i, P))\). Together with
the projected gradients \((\nabla_r E, \nabla_p E)\) from above, these form the
inputs to a Gaussian process regression with derivative observations
\cite{goswamiEfficientImplementationGaussian2025,mardiaKrigingSplinesDerivative1996,erikssonScalingGaussianProcess2018c}.
We construct the energy surface using the Inverse Multiquadric (IMQ) kernel:

\begin{equation}
k(\mathbf{x}, \mathbf{x}') = \left(c^2 + r^2\right)^{-1/2}, \quad r^2 = \|\mathbf{x} - \mathbf{x}'\|^2
\end{equation}

where \(c\) is a scale parameter. The function \(g(t) = (c^2 + t)^{-1/2}\) is
completely monotone on \([0, \infty)\): \((-1)^n g^{(n)}(t) \geq 0\) for all \(n
\geq 0\). By Schoenberg's theorem \cite{schoenbergMetricSpacesCompletely1938}, any kernel of the
form \(k = g(\|\mathbf{x} - \mathbf{x}'\|^2)\) with \(g\) completely monotone is
strictly positive definite on \(\mathbb{R}^d\) for all \(d\)
\cite{buhmannRadialBasisFunctions2003}. This guarantees non-singular interpolation matrices
regardless of point configuration.

The gradient-enhanced GP requires derivative kernel blocks. For the IMQ these
are:

\begin{align}
\frac{\partial k}{\partial x_i} &= -(x_i - x_i')\,(c^2 + r^2)^{-3/2} \\
\frac{\partial^2 k}{\partial x_i \partial x_j'} &= \delta_{ij}(c^2 + r^2)^{-3/2} - 3(x_i - x_i')(x_j - x_j')(c^2 + r^2)^{-5/2}
\end{align}

which assemble into the augmented kernel matrix
\cite{kochenderferAlgorithmsOptimization} over the observation vector
\(\mathbf{y}_{\text{full}} = [E_1, \nabla_r E_1, \nabla_p E_1, E_2, \ldots]\).
In practice, the derivative covariances are computed via automatic
differentiation using the JAX library, so extending to other kernels requires
only changing the base kernel function.

As \(r \to \infty\), the IMQ kernel decays as \(r^{-1}\), its first derivatives as
\(r^{-2}\), and second derivatives as \(r^{-3}\). This polynomial decay contrasts
with the exponential decay of the Matérn and squared exponential families. Heavy
tails allow the kernel to capture long-range basin structure from sparse path
data, and the augmented kernel matrix remains better conditioned than
exponentially decaying counterparts. Including derivative observations
effectively triples the information content per sampled geometry without
additional energy evaluations.

The hyperparameter optimization follows the subset optimization
\cite{goswamiAdaptivePruningIncreased2025b} concept from the OT-GPD; here the
points from the final path are used to calculate the length and noise scales,
which are subsequently applied while fitting the entire surface. This subsampled
approach reduces computational cost from \(O(N^3)\) to \(O(n^3)\) where \(n \ll N\)
(typically \(n \approx 20\) vs \(N \approx 500-2000\)), while simultaneously
improving robustness by isolating hyperparameter learning from transient
optimization dynamics.

The GP posterior naturally provides a pointwise variance estimate. We overlay
variance contours on the energy surface to delineate data-supported regions from
extrapolated ones. Near the sampled path, the posterior variance is small;
it grows with distance from the data and serves as a built-in reliability
indicator for the interpolation.

For systems producing more than approximately \(10^3\) gradient-augmented
observations (e.g., periodic slab calculations with many NEB steps), the
\(O(N^3)\) cost of the full GP becomes prohibitive. We use a Nystrom low-rank
approximation \cite{williamsUsingNystromMethod2000} adapted to the
gradient-enhanced setting. Given \(M \ll N\) inducing points
\(\{\mathbf{x}_m\}_{m=1}^M\) selected from the training set, the augmented
kernel matrix \(\mathbf{K}_{NN}\) (of size \(N(D+1) \times N(D+1)\), where \(D=2\)
is the input dimension) is approximated as

\begin{equation}
\tilde{\mathbf{K}}_{NN} \approx \mathbf{K}_{NM}\,\mathbf{K}_{MM}^{-1}\,\mathbf{K}_{MN}
\end{equation}

where \(\mathbf{K}_{NM}\) and \(\mathbf{K}_{MM}\) are the cross and inducing
kernel matrices, each containing the value, gradient, and Hessian blocks of the
IMQ kernel. The predictive mean becomes \(\hat{f}(\mathbf{x}_*) =
\mathbf{k}_{*M}\,\boldsymbol{\alpha}_M\) with weights obtained from

\begin{equation}
\boldsymbol{\alpha}_M = \mathbf{L}^{-\top}\left(\mathbf{S}^{-1}\,\mathbf{V}\,\mathbf{y}_{\text{full}}\right), \quad \mathbf{S} = \mathbf{V}\mathbf{V}^\top + \sigma^2\mathbf{I}_{M(D+1)}
\end{equation}

where \(\mathbf{L}\) is the Cholesky factor of \(\mathbf{K}_{MM} + \sigma^2
\mathbf{I}\) and \(\mathbf{V} = \mathbf{L}^{-1}\mathbf{K}_{MN}\). Inducing
points are selected from the last converged NEB path(s), which concentrate
observations near the final MEP geometry. This reduces the dominant cost from
\(O(N^3)\) to \(O(NM^2)\) (typically \(M = 300\), \(N > 5000\)) and makes the method
applicable to crystalline systems.

The resulting interpolated surface enables direct overlay of reference
structures (e.g., DFT-optimized saddle points) onto MLIP-generated landscapes.
One can then assess whether a potential captures qualitatively correct barrier
topology even when geometric displacements occur. This is impossible
with 1D energy profiles, where the reaction coordinate axis itself
depends on path geometry.
\section{Results}
\label{sec:org45ed584}
\begin{figure*}[t!]
    \centering
    \includegraphics[width=0.68\textwidth]{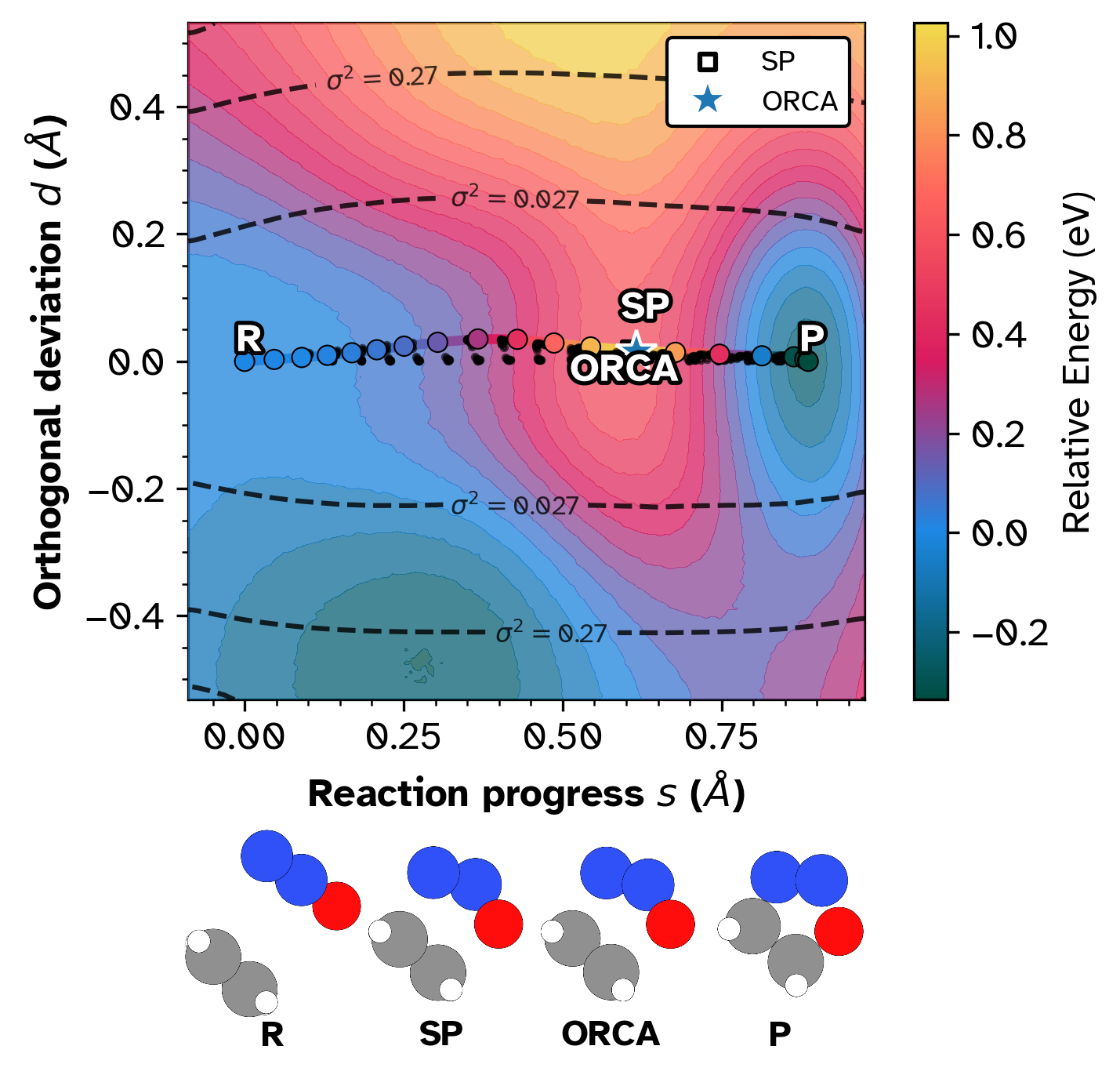}
    
    \vspace{0.3em}
    
    \includegraphics[width=0.32\textwidth]{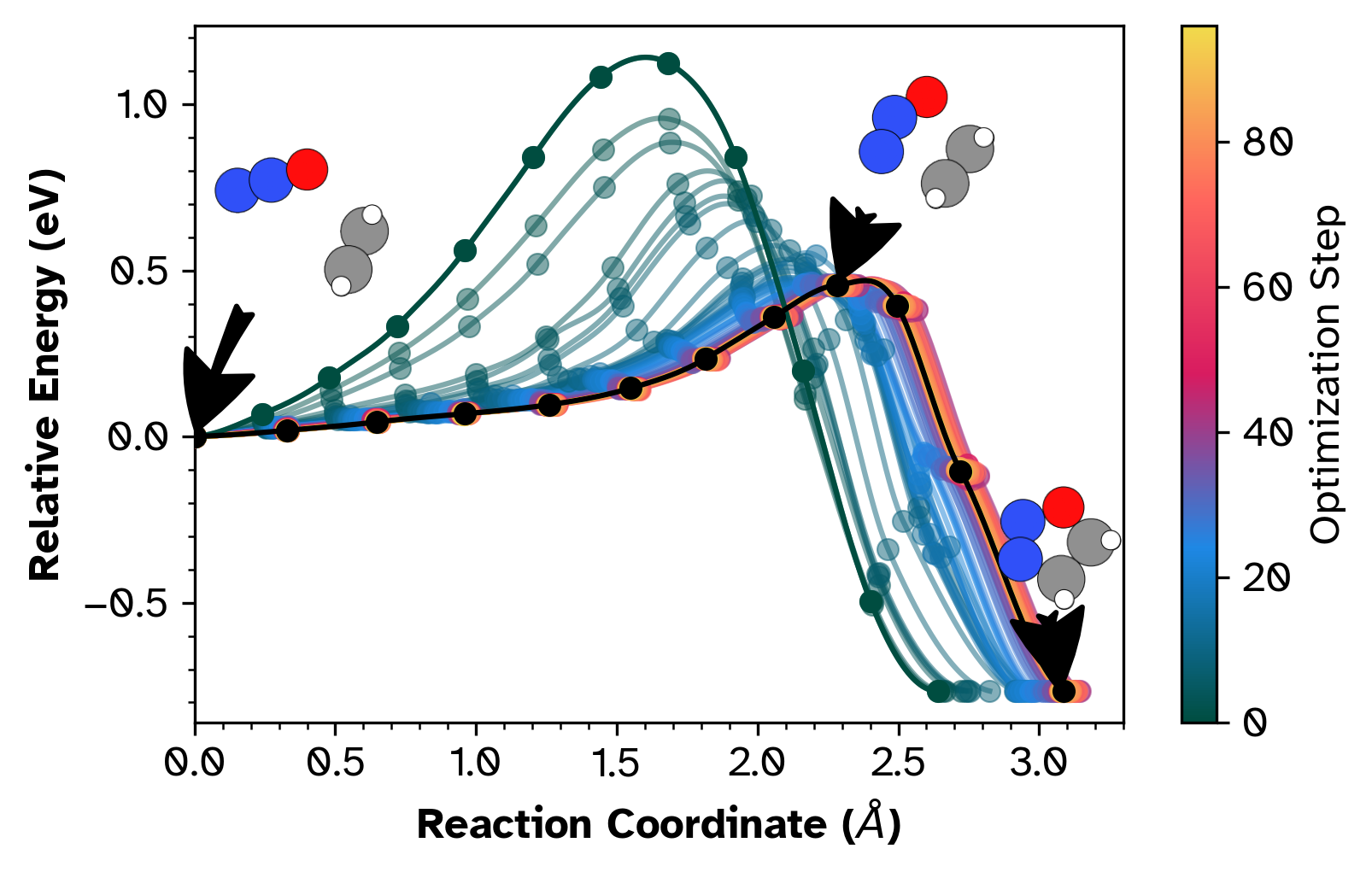}
    \hfill
    \includegraphics[width=0.32\textwidth]{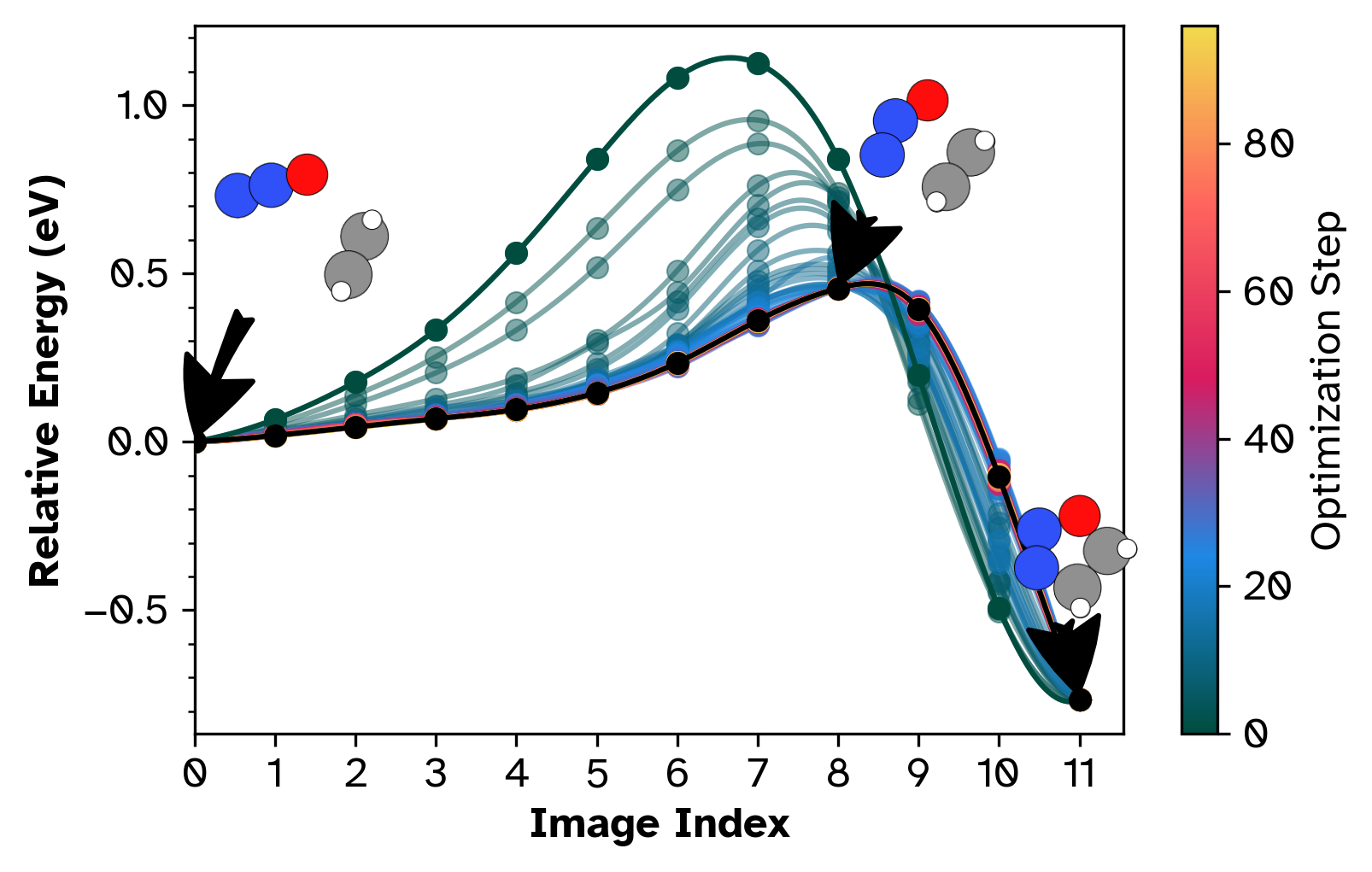}
    \hfill
    \includegraphics[width=0.32\textwidth]{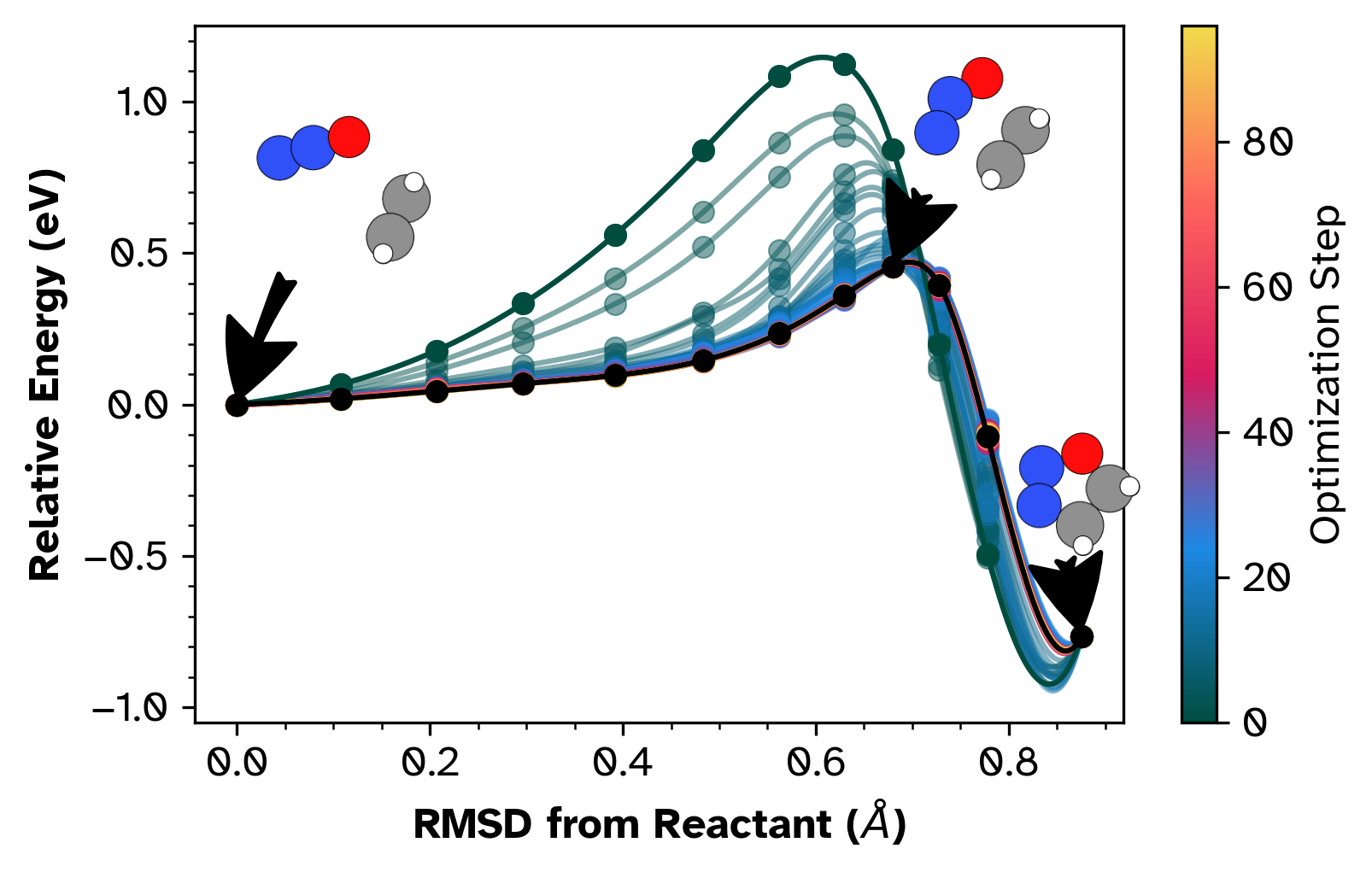}
    
    \caption{NEB optimization of ethylene + N\textsubscript{2}O cycloaddition
    using the MLIP. \textbf{Top:} 2D projection in $(s, d)$ coordinates
    (reaction progress vs.\ orthogonal distance) showing interpolated energy
    landscape (color), sampled structures (black dots), converged path (open
    circles), and posterior variance contours (dashed lines). White star
    indicates ORCA B3LYP-D3 saddle. \textbf{Bottom left:} Energy vs.\
    reaction coordinate. \textbf{Bottom center:} Energy vs.\ image index.
    \textbf{Bottom right:} Energy vs.\ RMSD from reactant. In all panels,
    colored curves show optimization progression (dark$\to$light =
    early$\to$late); final path in black.}
    \label{fig:assessment}
\end{figure*}

We empirically demonstrate the utility of the framework, first contrasting against 1D profiles for the 1,3-dipolar cycloaddition of ethylene and
N\textsubscript{2}O forming 4,5-dihydro-1,2,3-oxadiazole, a well-studied
benchmark reaction for NEB method development
\cite{koistinenNudgedElasticBand2019,asgeirssonNudgedElasticBand2021,bigiMetatensorMetatomicFoundational2025,marksIncorporationInternalCoordinates2025,ruttingerProtocolDirectingNudged2022,birkholzUsingBondingGuide2015}.
Figure \ref{fig:assessment} compares conventional one-dimensional representations with the
two-dimensional RMSD projection for energy-weighted NEB optimization in eOn \cite{rohitgoswamiTheochemUIEOnV29012026}
\footnote{from \url{https://eondocs.org}} using the PET-OMAT machine-learned potential
\cite{mazitovPETMADLightweightUniversal2025,bigiPushingLimitsUnconstrained2026} using Metatomic
\cite{bigiMetatensorMetatomicFoundational2025}. The model has been trained on
PBE reference data from the Open Materials dataset \cite{barroso-luqueOpenMaterials20242024}, with full construction details in the associated publication. For the cycloaddition, we contrast this with a saddle optimized from a different
NEB calculation using the B3LYP functional in ORCA
\cite{asgeirssonNudgedElasticBand2021} projected onto the interpolated landscape.

The conventional profiles (Figure \ref{fig:assessment}, bottom) show optimization from an initial barrier of approximately 1.1 eV to a final value near 0.4 eV over 120 steps, with the product lying approximately 0.8 eV below the reactant. The final path appears smooth and well-converged. These representations provide no information about sampling quality or the broader landscape topology.

The two-dimensional projection (Figure \ref{fig:assessment}, top) provides additional insight. Sampled structures (black dots) cluster tightly along the converged path, confirming robust single-pathway convergence. The interpolated energy surface displays a barrier region (yellow/orange, approximately 0.4 eV) separating the reactant basin from the deeper product basin (blue). The projection reveals that the ORCA saddle configuration overlaps with the estimate from the MLIP. The posterior variance contours (dashed lines) confirm that the interpolation is tightly constrained near the path and visibly degrades in the periphery, serving as a built-in reliability indicator for the surface estimate.

\begin{figure*}[t!]
    \centering
    \begin{subfigure}[b]{0.48\textwidth}
        \includegraphics[width=\textwidth]{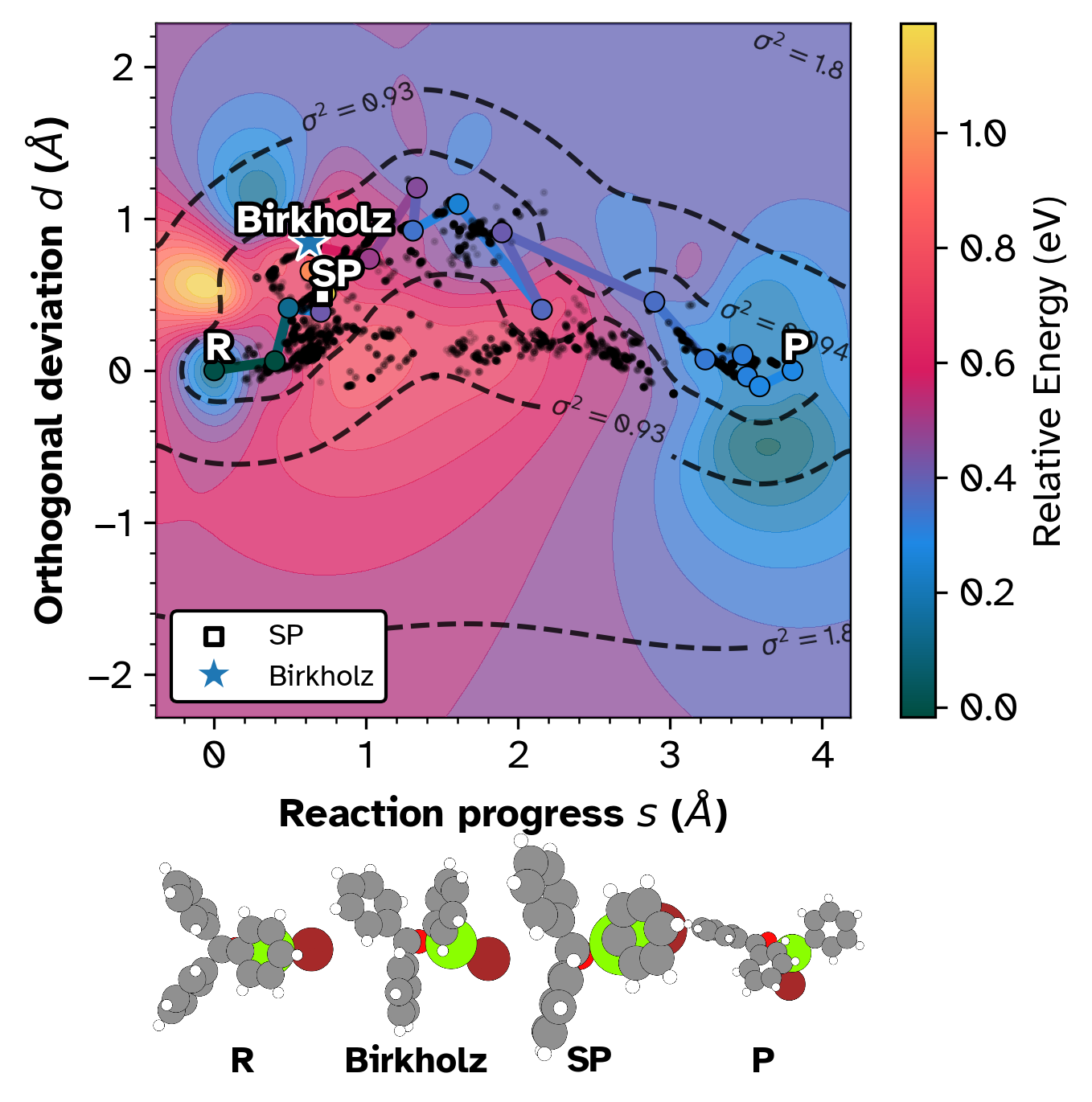}
        \caption{Grignard Rearrangement}
        \label{fig:grignard}
    \end{subfigure}
    \hfill
    \begin{subfigure}[b]{0.48\textwidth}
        \includegraphics[width=\textwidth]{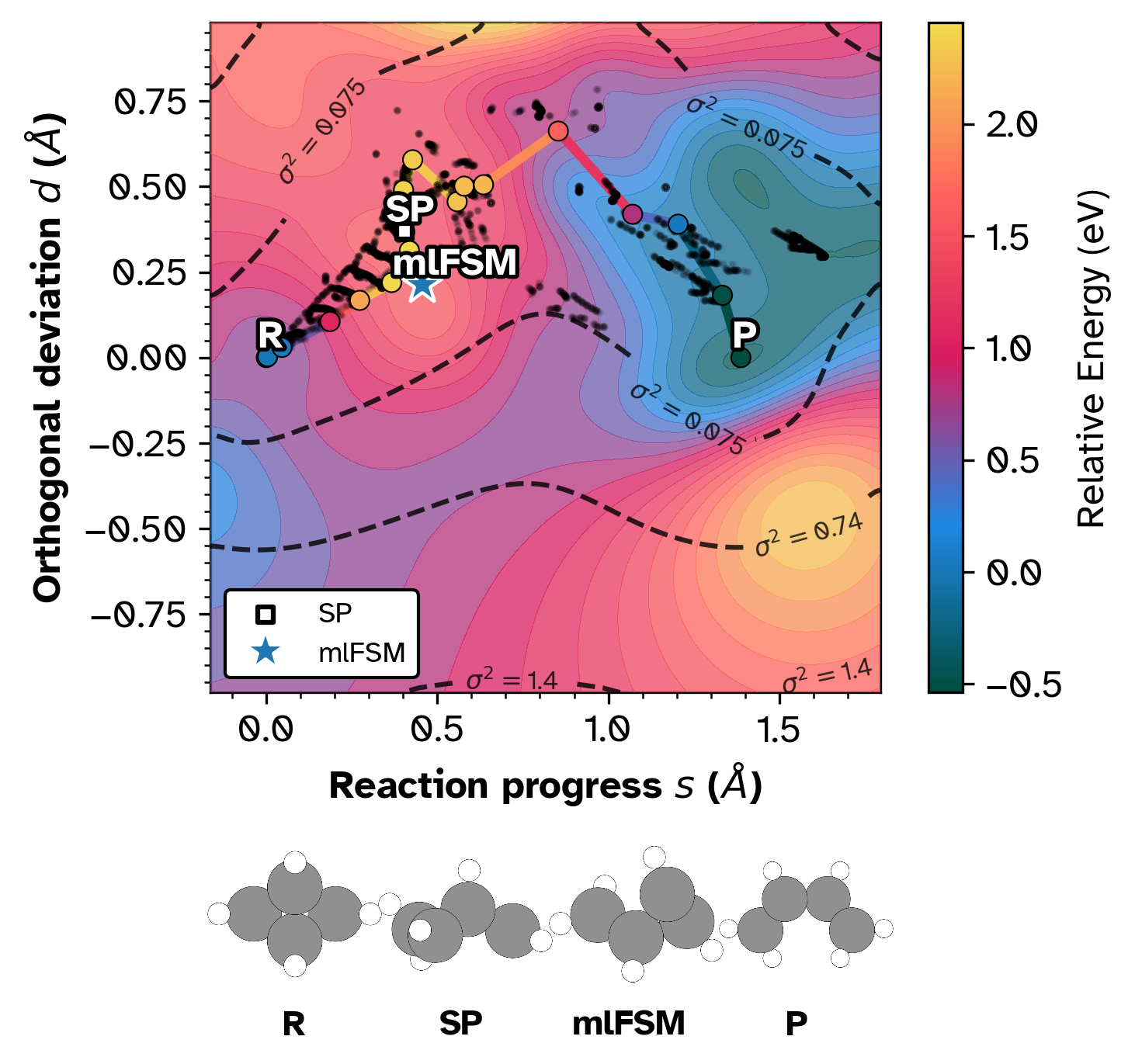}
        \caption{Conrotatory Bicyclobutane Ring Opening}
        \label{fig:bicyclo}
    \end{subfigure}
    
    \caption{Two-dimensional $(s, d)$ projections of complex rearrangement
    pathways with posterior variance contours (dashed lines). Both surfaces
    use the Nystr\"{o}m low-rank GP approximation; darker points denote the
    inducing subset (final converged NEB steps) while lighter points show
    the remaining optimization history retained for context.
    (a) The Grignard rearrangement of (Z)-2-phenyl-2-((trimethylsilyl)oxy)hex-4-enenitrile
    exhibits a distinct ``turning'' mechanism. The projection reveals optimization
    instability (scattered light dots) near the reactant basin that 1D profiles would obscure.
    The star indicates the reference saddle from Birkholz et al.
    (b) The conrotatory ring opening of bicyclobutane shows a sharp geometric ``kink''
    post-transition state. The projection confirms the machine-learned potential path
    (circles) faithfully traverses the saddle region identified by the Free String Method (star).}
    \label{fig:complex_cases}
\end{figure*}

To assess robustness on non-trivial topologies, we analyze the Grignard
rearrangement of (Z)-2-phenyl-2-((trimethylsilyl)oxy)hex-4-enenitrile
\cite{birkholzUsingBondingGuide2015} and the conrotatory ring opening of
bicyclobutane (Figure \ref{fig:complex_cases}) \cite{marksIncorporationInternalCoordinates2025}.
Unlike the linear character of the cycloaddition, these reactions proceed
through significant geometric turns. The Grignard pathway (Figure \ref{fig:grignard}) exhibits a
curved trajectory where the path tangent rotates nearly 90 degrees in the
\((s, d)\) plane. The 2D projection captures the optimization noise (scattered
lighter points) near the reactant basin, a diagnostic feature
lost in scalar reaction coordinate plots, while showing without explicit
structural inspection that the reference and computed saddles lie on the same
energy contour. Similarly, the conrotatory bicyclobutane ring opening (Figure \ref{fig:bicyclo})
displays a sharp geometric ``elbow'' immediately following the transition state.
The interpolated surface correctly places the reference Free String Method
(mlFSM) saddle \cite{marksIncorporationInternalCoordinates2025} within the
barrier region of the MLIP. This confirms that the NEB trajectory found the
physically relevant saddle despite the complex topology, a validation that
remains ambiguous when observing only the image index or scalar reaction
coordinate \footnote{or in practice, would involve normal mode analysis, barrier heights,
RMSD, and visual inspection}.

The 2D projection provides information beyond standard scalar comparisons.
Computing energy differences and RMSD values between two saddle
points tells one \emph{that} they differ, but not \emph{how} they differ relative to the
reaction landscape. A 1D energy profile, by construction, collapses all
geometric information onto the cumulative displacement axis. It cannot
distinguish whether two saddle estimates from different methods
lie on the same energy contour, sit in different basins, or whether the
optimization explored alternative pathways before converging. The 2D projection
preserves these geometric relationships. One can immediately see, for example,
that the ORCA and MLIP saddles in Figure \ref{fig:assessment} occupy the same barrier region
despite different path histories, or that the optimizer noise in the Grignard
case (Figure \ref{fig:grignard}) is confined to a specific geometric region rather than
distributed along the path. These are qualitative assessments, analogous to
plotting a trajectory on a map rather than reporting only the total distance
traveled.

The framework also provides a unified coordinate system for validating
potential energy surfaces against higher-level theory. If these disagree, the
relative differences between the saddles are apparent. The projections reveal
whether resulting structures sit within comparable energy contours. Determining
whether the energy surface captures the qualitative barrier topology, even when
the precise saddle geometry differs due to functional sensitivity, is impossible
with conventional one-dimensional profiles.

For complete reproduction, the landscape figures are generated by a single
command-line invocation. As an example, the Grignard projection (Figure \ref{fig:grignard}) is
produced by:

\begin{lstlisting}[language=bash,basicstyle=\ttfamily\scriptsize,breaklines=true,columns=fullflexible]
python -m rgpycrumbs.cli --dev eon plt-neb \
  --con-file neb.con --output-file jax_grad_imq.png \
  --plot-type landscape --project-path \
  --show-pts --landscape-path all \
  --plot-structures crit_points \
  --surface-type grad_imq_ny \
  --ira-kmax 14 --show-legend \
  --additional-con ts.xyz "Birkholz"
\end{lstlisting}

The \texttt{-{}-{}project-path} flag (on by default) rotates the raw \((r, p)\) RMSD
coordinates into the \((s, d)\) frame described above, \texttt{-{}-{}surface-type
grad\_imq\_ny} selects the Nystrom-accelerated gradient-enhanced IMQ kernel,
\texttt{-{}-{}ira-kmax} sets the adjacency cutoff distance for IRA graph-based structure
matching, \texttt{-{}-{}plot-structures crit\_points} renders the reactant, product,
saddle estimate, and highest-energy image, and \texttt{-{}-{}additional-con} overlays a
reference saddle from an external file. Cosmetic options (figure size, DPI, font
size, rotation) are omitted for brevity. The materials archive includes exact
commands and outputs at the time of submission.
\section*{Limitations}
\label{sec:limitations}
The \((r, p)\) projection is a lossy map from \(\mathbb{R}^{3N}\) to
\(\mathbb{R}^2\). Multiple distinct Cartesian configurations can share the same
RMSD coordinates, so the interpolated energy surface does not reconstruct the
true PES but rather a projected slice of it. Apparent features far from the
sampled data, for instance, alternative low-energy pathways visible in the color
map, may not correspond to physical pathways in the full configuration space.
The interpolation should be read as a qualitative guide near the data, not as a
quantitative surrogate for the potential. The posterior variance contours
overlaid on the figures directly visualize this degradation so that users can
distinguish data-supported features from extrapolation artifacts.

As discussed in the methodology, only the tangential force component enters the
gradient projection. The method does not recover curvature information orthogonal
to the path, and the MEP orthogonality property cannot be enforced in the 2D
projection. Any extension to orthogonal force projection would require
differentiable structure matching. For molecular systems where the permutation
group is large, this remains an open problem. For crystalline systems, where
space group symmetry constrains the permutation set, tractable approximations
may be feasible and are a natural direction for future work.

The choice of IMQ kernel, while theoretically motivated by its positive
definiteness and polynomial tail properties, is one of several valid options.
The software provides squared exponential, Matern 5/2, and thin plate spline
alternatives. Different kernels may be better suited to different reaction
topologies, and no single kernel is universally optimal.

Finally, the method is designed for post-hoc visualization of existing
calculations. It does not accelerate the NEB computation itself, nor does it
replace quantitative analysis (normal mode analysis, barrier heights, vibrational
frequencies) required for kinetic rate calculations.
\section*{Conclusion}
\label{sec:conclusion}
We presented a coordinate-free visualization method for analyzing reaction path
optimization trajectories. By projecting high-dimensional pathways onto a
surface defined by permutation-invariant RMSD coordinates, the method reveals
geometric and energetic features obscured by standard one-dimensional profiles.

In current practice, comparing NEB results across methods or potentials requires
either pointwise structural inspection (impractical beyond a handful of images)
or scalar summaries (barrier height, RMSD to reference) that discard geometric
context. The 2D projection occupies the space between these extremes. A single
plot preserves the geometric relationship between all sampled structures, the
converged path, and any reference points, while remaining interpretable at a
glance. The benchmarks show that this representation distinguishes numerical
artifacts from physical topology and reveals whether different potentials produce
qualitatively similar barrier regions, even when the precise saddle geometries
differ. The rotated \((s, d)\) frame separates reaction progress from path
deviation and gives a more physical coordinate system than raw RMSD distances.
The posterior variance contours supply built-in reliability assessment for
the interpolated surface. A Nystrom low-rank approximation extends the method
to crystalline systems with large numbers of observations.

While we focus on double-ended methods here, the projection operates
independently of the path generation algorithm. One can project histories from
single-ended saddle search methods, molecular dynamics, or metadynamics onto
these intrinsic axes post hoc to diagnose path quality and basin hysteresis.
The relative energy surfaces from multiple machine learning models could also be
compared within such a framework.
\section*{Acknowledgments}
\label{sec:acknowledgments}
The author thanks Michele Ceriotti for suggesting the connection to the path
collective variable formalism, and the members of Lab-COSMO at EPFL for
discussions. The author thanks the anonymous reviewers for substantially
improving the manuscript and the editor for facilitating the discussion.
The author also thanks his family, pets, and plants for their patience and
support. The author acknowledges the color scheme designed by Ruhila Goswami
used for the figures. 
\section*{Conflict of Interest}
\label{sec:coi}
We declare no conflicts of interest.
\bibliography{nebViz2025}
\end{document}

%% file: authors_arxiv.tex
\author{ \href{https://orcid.org/0000-0002-2393-8056}{\includegraphics[scale=0.06]{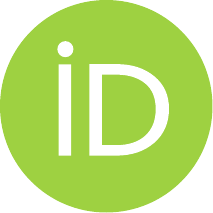}\hspace{1mm}Rohit Goswami}\thanks{Corresponding Author} \\
  Institute IMX and Lab-COSMO \\
  \'Ecole polytechnique f\'ed\'erale de Lausanne (EPFL) \\
  Station 12, CH-1015 Lausanne, Switzerland \\
  and \\
  TurtleTech ehf., 107 Reykjav\'ik, Iceland\\[1ex]
  \texttt{rgoswami@ieee.org}
}